\documentclass[twocolumn,aps,prl,amsmath,groupedaddress]{revtex4}
\usepackage[T1]{fontenc}
\usepackage[latin9]{inputenc}
\usepackage{amsmath}
\usepackage{amssymb}
\usepackage{graphicx}

\makeatletter
\@ifundefined{textcolor}{}
{%
 \definecolor{BLACK}{gray}{0}
 \definecolor{WHITE}{gray}{1}
 \definecolor{RED}{rgb}{1,0,0}
 \definecolor{GREEN}{rgb}{0,1,0}
 \definecolor{BLUE}{rgb}{0,0,1}
 \definecolor{CYAN}{cmyk}{1,0,0,0}
 \definecolor{MAGENTA}{cmyk}{0,1,0,0}
 \definecolor{YELLOW}{cmyk}{0,0,1,0}
}

\preprint{}

\makeatother

\begin{document}

\title{Qubit architecture with high coherence and fast tunable coupling}

\author{Yu Chen$^{1}$}

\thanks{These authors contributed equally to this work}

\author{C. Neill$^{1}$}

\thanks{These authors contributed equally to this work}

\author{P. Roushan$^{1}$}

\thanks{These authors contributed equally to this work}

\author{N. Leung$^{1}$}

\author{M. Fang$^{1}$}

\author{R. Barends$^{1}$}

\author{J. Kelly$^{1}$}

\author{B. Campbell$^{1}$}

\author{Z. Chen$^{1}$}

\author{B. Chiaro$^{1}$}

\author{A. Dunsworth$^{1}$}

\author{E. Jeffrey$^{1}$}

\author{A. Megrant$^{1}$}

\author{J. Y. Mutus$^{1}$}

\author{P. J. J. O'Malley$^{1}$}

\author{C. M. Quintana$^{1}$}

\author{D. Sank$^{1}$}

\author{A. Vainsencher$^{1}$}

\author{J. Wenner$^{1}$}

\author{T. C. White$^{1}$}

\author{Michael R. Geller$^{2}$}

\author{A. N. Cleland$^{1}$}

\author{John M. Martinis$^{1}$}

\email{martinis@physics.ucsb.edu}

\affiliation{$^{1}$Department of Physics, University of California, Santa Barbara,
California 93106-9530, USA}

\affiliation{$^{2}$Department of Physics and Astronomy, University of Georgia,
Athens, Georgia 30602, USA}
\begin{abstract}
We introduce a superconducting qubit architecture that combines high-coherence
qubits and tunable qubit-qubit coupling. With the ability to set the
coupling to zero, we demonstrate that this architecture is protected
from the frequency crowding problems that arise from fixed coupling.
More importantly, the coupling can be tuned dynamically with nanosecond
resolution, making this architecture a versatile platform with applications
ranging from quantum logic gates to quantum simulation. We illustrate
the advantages of dynamic coupling by implementing a novel adiabatic
controlled-Z gate, at a speed approaching that of single-qubit gates.
Integrating coherence and scalable control, our ``gmon'' architecture
is a promising path towards large-scale quantum computation and simulation. 
\end{abstract}
\maketitle
The fundamental challenge for quantum computation and simulation is
to construct a large-scale network of highly connected coherent qubits
\cite{Nielsen2000,You2005}. Superconducting qubits use macroscopic
circuits to process quantum information and are a promising candidate
towards this end \cite{Devoret2013}. Over the last several years,
materials research and circuit optimization have led to significant
progress in qubit coherence \cite{Megrant2012,Barends2013,Paik2011}. Superconducting
qubits can now perform hundreds of operations within their coherence
times, allowing for research into complex algorithms such as error
correction \cite{Barends2014,Chow2013}.

It is desirable to combine these high-coherence qubits with tunable
inter-qubit coupling; the resulting architecture would allow for both
coherent local operations and dynamically varying qubit interactions.
For quantum simulation, this would provide a unique opportunity to
investigate dynamic processes in non-equilibrium condensed matter
phenomena \cite{Buluta2009,Houck2012,Geller2012,Polkovnikov2011,Calabrese2012}.
For quantum computation, such an architecture would provide isolation
for single-qubit gates while at the same time enabling fast two-qubit
gates that minimize errors from decoherence. Despite previous successful
demonstrations of tunable coupling \cite{Allman2010,Bialczak2011,Harris2007,Hime2006,Niskanen2007,Ploeg2007,Srinivasan2011,Blais2003,Liu2006,Pinto2010},
these applications have yet to be realized due to the challenge of
incorporating tunable coupling with high coherence devices.

\begin{figure}[b!]
\includegraphics{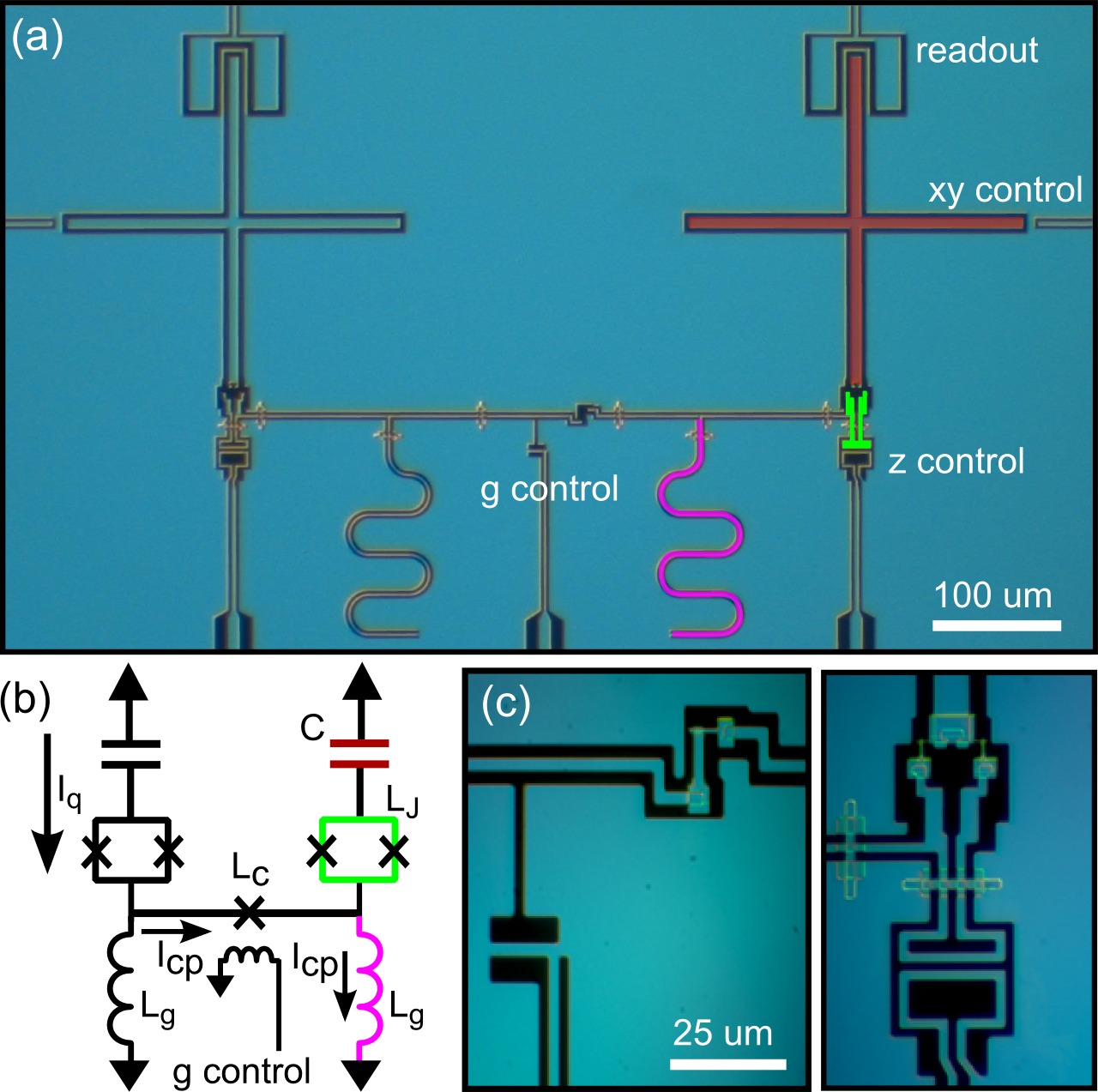} \caption{(a) Optical micrograph of two inductively coupled gmon qubits. The
cross-shaped capacitors are placed in series with a tunable Josephson
junction and followed by a linear inductor to ground. The circuit
is depicted schematically in (b) with arrows indicating the flow of
current for an excitation in the left qubit. The qubits are connected
with a line containing a junction that acts as a tunable inductor
to control the coupling strength. (c) Micrographs of the coupler junction
(left) and qubit SQUID (right). The bottom of each image shows a bias
line used to adjust the coupling strength (left) and qubit frequency
(right, not shown in schematic). }

\label{fig:device} 
\end{figure}

Here, we introduce a planar qubit architecture that combines high
coherence with tunable inter-qubit coupling $g$. This ``gmon''
device is based on the Xmon transmon design \cite{Barends2013}, but
now gives nanosecond control of the coupling strength with a measured
on/off coupling ratio exceeding 1000. We find that our device retains
the high coherence inherent in the Xmon design, with the coupler providing
unique advantages in constructing single- and two-qubit quantum logic
gates. With the coupling turned off, we demonstrate that our architecture
is protected from the frequency crowding problems that arise from
fixed coupling. Our single-qubit gate fidelity is nearly independent
of the qubit-qubit detuning, even when operating the qubits on resonance.
By dynamically tuning the coupling, we implement a novel adiabatic
controlled-Z gate at a speed approaching that of single-qubit gates.

\begin{figure}[t!]
\includegraphics{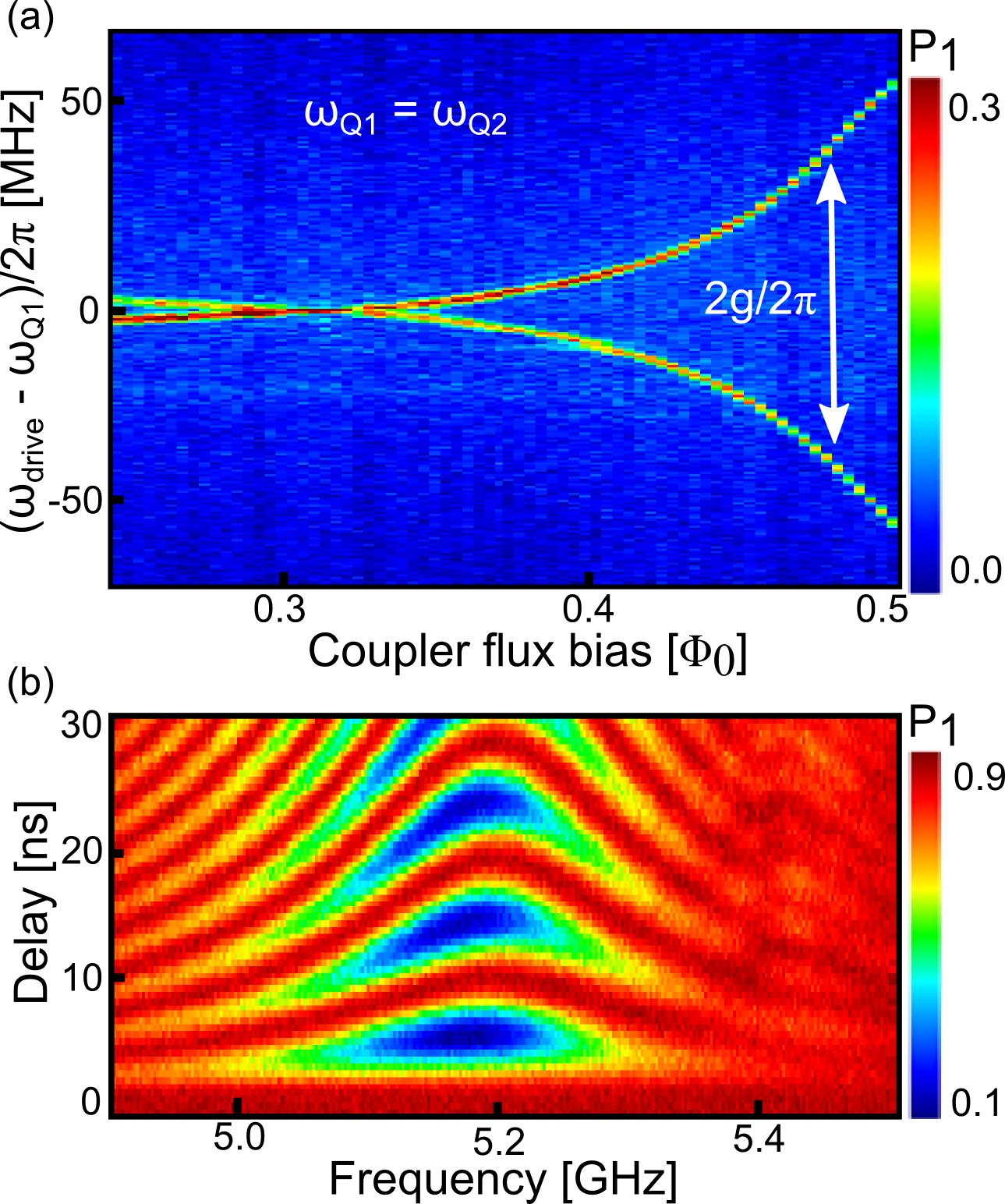} \caption{(a) The dependence of the coupling strength on the coupler flux bias
while the two qubits are on resonance, with $\omega_{Q1}/2\pi=\omega_{Q2}/2\pi=5.67\,$GHz.
For each value of the coupler flux bias, we sweep the microwave drive
frequency and measure the excited state probability $P_{1}$ (colorbar)
of $Q_{1}$. There are two distinct peaks in the spectroscopy that
result from an energy level splitting. The frequency splitting is
twice the coupling strength $g/2\pi$ and ranges from 0 to 110 MHz.
(b) $Q_{1}$ excited state probability (colorbar) versus the frequency
of $Q_{1}$ (horizontal axis) after exciting the qubit and waiting
a variable delay time (vertical axis). $Q_{2}$ is fixed at 5.18 GHz
and the coupling is set to 55 MHz. On resonance, the two qubits swap
an excitation in 5 ns. }

\label{fig:coupling} 
\end{figure}

%\section{2. Device - Basics}

A two-qubit unit cell with tunable coupling is shown in Fig.\,\ref{fig:device}(a).
The qubits and control lines are defined by an aluminum film with
cuts exposing the underlying sapphire substrate. Our circuit design
is based on the Xmon qubit, consisting of a cross-shaped capacitor
resonating with a nonlinear inductor $L_{J}=9.0\,\textrm{nH}$ made
from a SQUID. We modify the Xmon design to introduce a linear inductor
$L_{g}=200\,\textrm{pH}$ from the junction to ground, with $L_{g}\ll L_{J}$
so that the qubit nonlinearity is largely unaffected (see Ref.\,\cite{Supplement}
for a detailed discussion). This inductor introduces a node in the
circuit where current from one qubit can be tapped off to interact
with a neighboring qubit. A junction connecting the two nodes acts
as a tunable inductance $L_{c}$ that controls the flow of this current
and therefore the coupling.

%\section{3. Device - Theory}

The physics behind this tunable coupler is well explained using a
simple linear model, since the coupling currents are much smaller
than the critical current of the coupling junction $I_{0}=330\,\textrm{nA}$;
see Ref.\,\cite{Geller2014} for a full quantum mechanical treatment.
A circuit diagram for the device is given in Fig.\,\ref{fig:device}(b).
An excitation current in the first qubit $I_{q}$ mostly flows through
$L_{g}$, with a small fraction $I_{\textrm{cp}}=I_{q}L_{g}/(2L_{g}+L_{c})$
flowing through the coupler to the second qubit. This current generates
a flux in the second qubit $\Phi_{2}=L_{g}I_{\textrm{cp}}$. 
In the absence of parasitic inductance, the effective mutual inductance can be expressed as 
\begin{align}
M=\frac{\Phi_{2}}{I_{q}}=\frac{L_{g}^{2}}{2L_{g}+L_{c}}\ \ .\label{eq:M}
\end{align}

Using this mutual inductance, the interaction Hamiltonian for the
two qubits on resonance can be written as 
\begin{align}
\hat{H}_{int}=-\frac{\omega_{0}}{2}\frac{M}{L_{J}+L_{g}}(\hat{a}_{1}^{\dagger}\hat{a}_{2}+\hat{a}_{1}\hat{a}_{2}^{\dagger})\ \ ,\label{eq:H}
\end{align}
where $\omega_{0}$ is the qubit resonance frequency. This equation
uses the rotating wave approximation to express photon swapping via
the raising and lowering operators \cite{Pinto2010}. The coefficient
of the interaction Hamiltonian gives the coupling strength 
\begin{align}
g=-\frac{\omega_{0}}{2}\frac{L_{g}}{L_{J}+L_{g}}\frac{L_{g}}{2L_{g}+L_{c0}/\cos\delta}\ \ ,\label{eq:g}
\end{align}
where $L_{c}$ is replaced by the Josephson inductance $L_{c}=\Phi_{0}/(2\pi I_{0}\cos\delta)\equiv L_{c0}/\cos\delta$.
Here $\delta$ is the phase difference across the coupler junction,
set by applying a DC flux. Note that the DC current from this flux
flows only through the coupler, not through the qubit junctions because
of their series capacitance. The coupling $g$ can be varied continuously
from negative to positive, going smoothly through zero at $\delta=\pi/2$.
This smooth transition from positive to negative ensures the existence
of a bias such that the coupling is completely negated; this is true
even in the presence of small stray coupling.

%\section{4. Device - Discussion}

A critical part of our design is the compatibility between high coherence
and tunable coupling. The key concept in maintaining coherence is
the voltage divider created by $L_{J}$ and $L_{g}$: placing the
coupling circuit at this low voltage node reduces capacitive losses
by a factor of $(L_{J}/L_{g})^{2}$ -- over 2000 in our design. For
the gmon, we measure an energy relaxation time $T_{1}$ $\sim$ 7--10\,$\mu$s and is independent of the coupling strength (see Ref.\,\cite{Supplement}).
This is comparable to the performance of previous Xmon devices with
similar capacitor geometry (8$\,\mu$m center trace, 4$\,\mu$m gap)
and aluminum deposition conditions (high vacuum e-beam evaporation).
Devices grown with molecular beam epitaxy and with optimized capacitor
geometry have been shown to have lifetimes exceeding 40 $\mu$s \cite{Barends2013}.

%\section{5. Core performance}

The core functionality of the gmon coupler is demonstrated in Fig.\,\ref{fig:coupling}.
In panel (a) we show the variation of the coupling strength as a function
of the coupler flux bias, for the condition where the two qubits are
brought into resonance at a frequency $\omega_{0}/2\pi=5.67\,\textrm{GHz}$.
Here for one qubit we sweep the microwave drive frequency and measure
the qubit excited state probability $P_{1}$. We observe two distinct
resonances at frequencies $\omega_{0}+g$ and $\omega_{0}-g$ that
result from the coupling-induced energy level splitting. The total
splitting is twice the coupling strength, ranging from 0 to 110 MHz.
This range can be further increased by modifying the critical current of the coupler junction. 
Note that we have compensated for
the small changes in the qubit frequency ($\sim g$) that occur as
$L_{c}$ is varied;Ref.\,\cite{Supplement} gives details on how
these controls are effectively made orthogonal.

\begin{figure}[t!]
\includegraphics{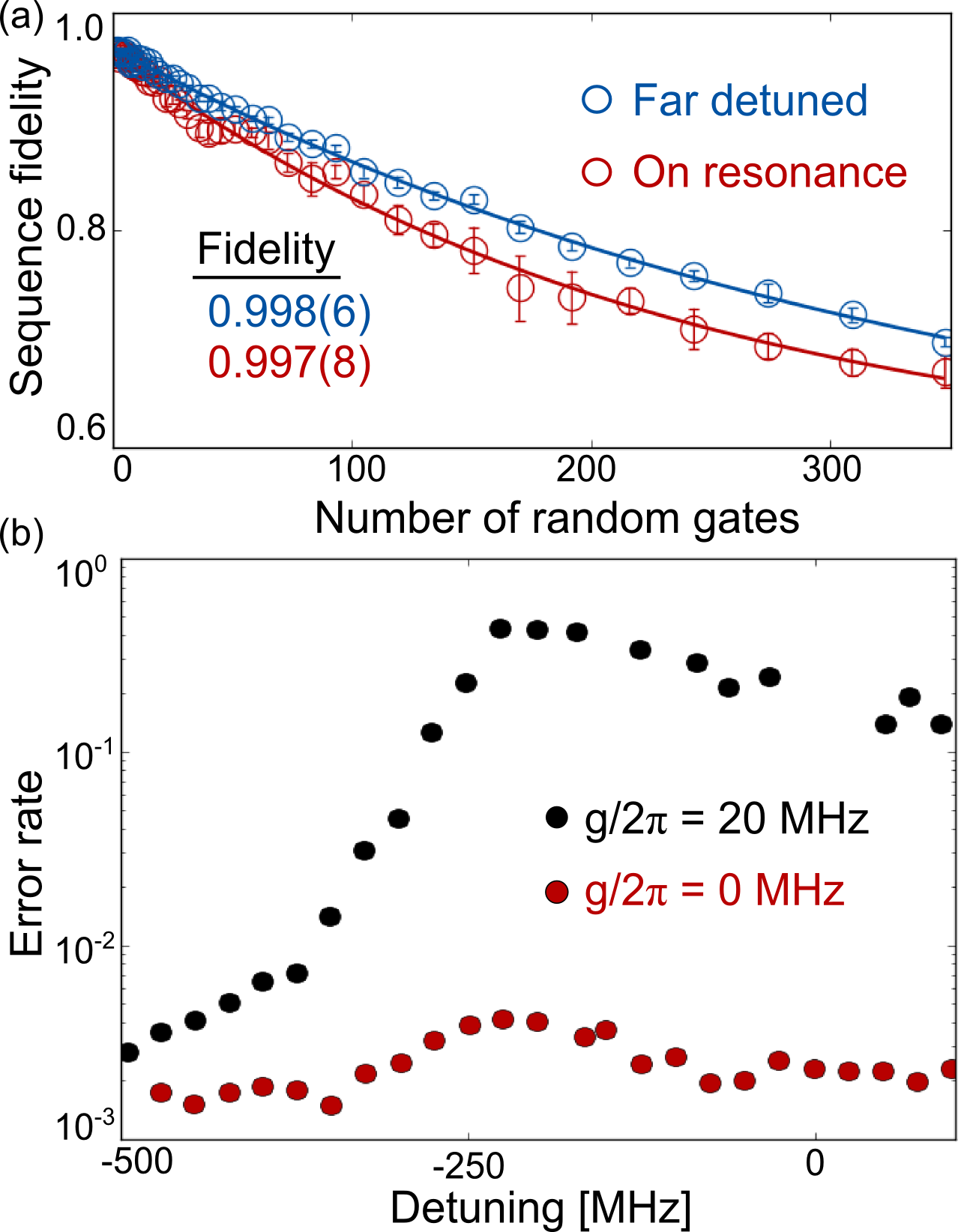} \caption{ Simultaneous single-qubit randomized benchmarking. (a) The raw benchmarking
data for $Q_{1}$ when $Q_{2}$ is far detuned (blue) and on resonance
with random gates applied to both qubits (red). Operating the qubits
on resonance degrades the gate performance by $<.1\%$. Lines are
fits to a decaying exponential. (b) The average error rate for $Q_{1}$
as a function of the detuning between the two qubits, shown in red
for nominally zero coupling and in black for 20 MHz coupling. The
ability to turn off the coupling results in an error rate that is
nearly flat, with a value on resonance that is two orders of magnitude
lower than for moderate fixed coupling.}

\label{fig:singleQubit} 
\end{figure}

For the data in Fig.\,\ref{fig:coupling}(b) we set the the coupling
strength to its maximum value and rapidly exchange an excitation between
the two qubits. We excite the first qubit ($Q_{1}$), turn on the
coupling, wait a variable delay time, and then measure the excited
state probability of $Q_{1}$. We vary the frequency of $Q_{1}$ while
fixing that of the second qubit ($Q_{2}$). The resonance interaction
results in the expected chevron pattern \cite{Hofheinz2009}. The strong coupling allows
the excitation to swap between the two qubits in 5\,ns, consistent
with the 110 MHz splitting measured above. At this rate, a $\sqrt{i\mbox{SWAP}}$
gate could generate a Bell state in 2.5\,ns, whereas a non-adiabatic
CZ could be implemented in 10\,ns \cite{Yamamoto2010}. We have also
performed the same measurement with nominally zero coupling (seeRef.\,\cite{Supplement})
and observe no indication of swapping after 6\,$\mu$s. This places
an upper bound on the residual coupling of 50\,kHz, providing an
on/off ratio of over 1000.

%\section{7.  Randomized Benchmarking - Theory and introduction}

By incorporating tunable coupling with high coherence qubits, our
architecture provides a viable platform for both quantum computation
and simulation. We have applied this device to quantum simulation
in a separate experiment where we have demonstrated an interaction-driven
topological phase transition \cite{Roushan2014}. In the following,
we focus on applications in quantum computation by implementing elementary
logic gates. This architecture offers two distinct advantages: the
ability to decouple qubits for local single-qubit gates and the ability
to dynamically tune the interaction for fast two-qubit gates.

We characterize gate performance using a simplified form of randomized
benchmarking \cite{Knill2008,Magesan2012}, which applies a series
of Pauli gates that move the qubit among the 6 states $\{|0\rangle,|1\rangle,(|0\rangle+|1\rangle)/\sqrt{2},(|0\rangle+i|1\rangle)/\sqrt{2},(|0\rangle-|1\rangle)/\sqrt{2},(|0\rangle-i|1\rangle)/\sqrt{2}\}$.
These gates belong to a subset of the Clifford group and are generated
using microwave pulses that correspond to Bloch sphere rotations of
angle $\pi$ and $\pi/2$ around the X and Y axis. From this set we
randomly choose $m$ gates and apply these to the qubit, including
a final gate that ideally maps the qubit back into the ground state. The
probability of finding the qubit in the ground state is called the
sequence fidelity $F_{\textrm{seq}}$, which decays exponentially
with the number of gates by $F_{\textrm{seq}}=Ap^{m}+B$. Here $A$,
$B$ and $p$ are fit parameters; $A$ and $B$ relate to state preparation
and measurement. We are interested in the average error per gate $r$,
determined through the relation $r=(1-p)(d-1)/d$ where $d=2^{N_{\textrm{qubits}}}$.
We note that Pauli gates do not fully depolarize errors, hence the
extracted gate fidelities are only indicative.

%\section{8.  Randomized Benchmarking - Experimental data}
The ability to isolate individual qubits for local operations is one
advantage offered by a tunable coupling architecture. A metric to
quantify this isolation is single-qubit gate fidelity $1-r$. As a
baseline, we perform randomized benchmarking on the first qubit while
the second qubit is far detuned and effectively decoupled. The sequence
fidelity is plotted in Fig.\,\ref{fig:singleQubit}(a) and displays
the expected exponential decay with the number of random gates. Fitting
the decay curve yields an average single-qubit gate fidelity of $99.86\%$.
The two qubits are then placed on resonance with $g=0$ and the measurement
is repeated on both qubits; data for the first qubit is shown. Simultaneously
operating the two qubits on resonance reduces the gate fidelity by
$<0.1\%$. The added error results from two sources: residual inter-qubit
coupling and imperfect cancellation of microwave crosstalk between
the control signals.

\begin{figure}[t!]
\includegraphics{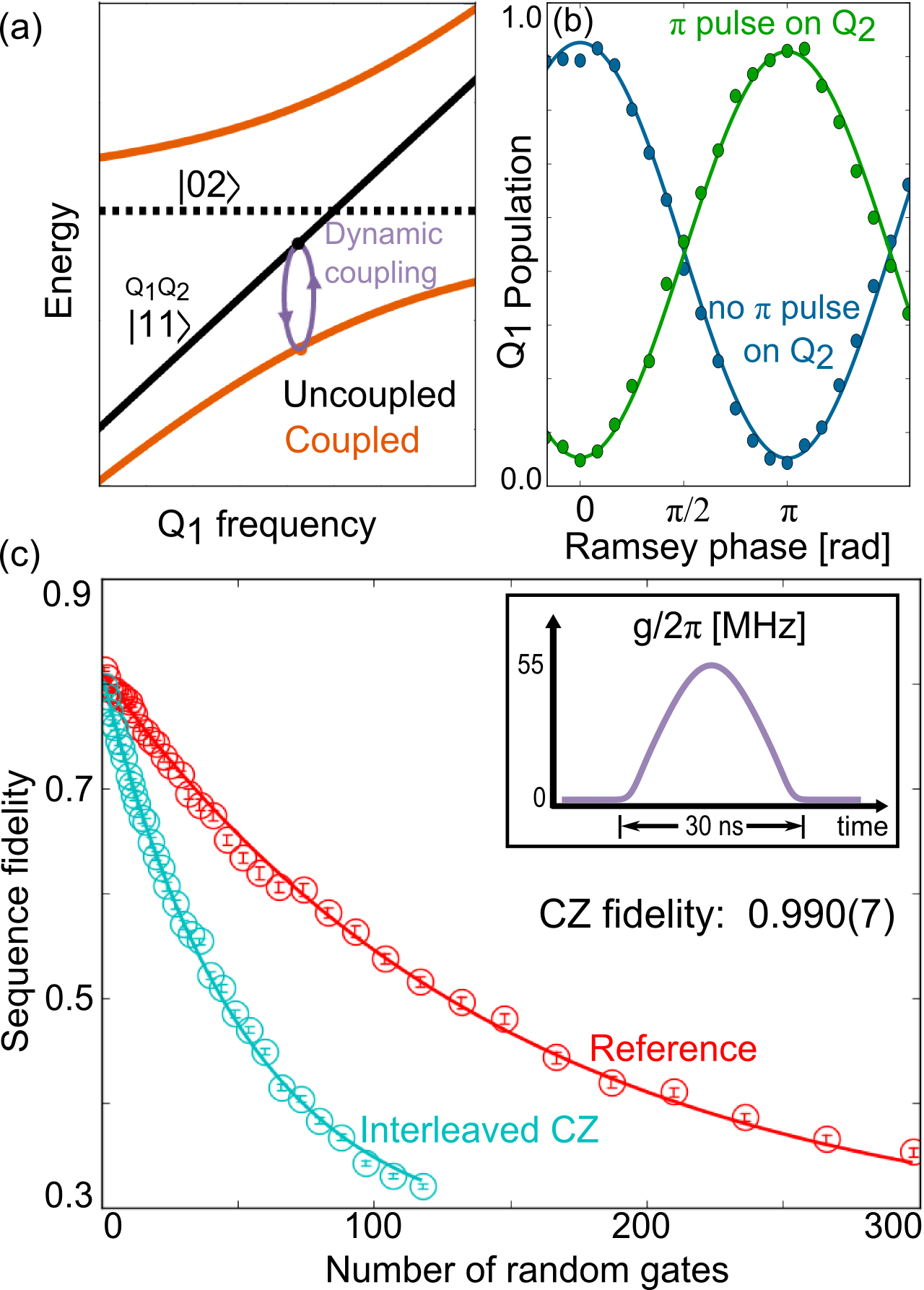} \caption{ (a) Energy level diagram; illustration of a CZ using tunable coupling.
Black lines are the uncoupled two-photon eigenenergies; orange lines
are the coupled eigenenergies. As the coupling is tuned on and off
(depicted in purple), the energy levels repel and the states accumulate
a dynamic phase. (b) Ramsey data demonstrating zero phase shift for
single-photon states and a $\pi$ phase shift for the two-photon state.
(c) Randomized benchmarking results for a CZ gate utilizing the pulse
shape shown inset. We are able to achieve 99.07\% fidelity with a
30 ns gate. }

\label{fig:twoQubits} 
\end{figure}

In panel (b), we repeat this measurement as a function of the frequency
separation of the two qubits, demonstrating the effects of frequency
crowding that result from fixed coupling. The average error rate is
plotted in Fig.\,\ref{fig:singleQubit}(b) for both $g/2\pi=0$ and
20\,MHz; note that the latter value is only a third of the maximum
possible coupling. Even for this relatively weak interaction, the
single-qubit gate fidelity undergoes a significant reduction for detuning
less than 500\,MHz. 
The ability to turn off the coupling $g$ results
in an error rate that is nearly flat, with a value on resonance that
is two orders of magnitude lower than for fixed coupling. We note
there is a slight degradation in the qubit performance near the qubit
nonlinearity (220 MHz).

A concern in designing transmon has been the cross coupling of qubits.
One approach to resolve this has been the use of 3D devices in
which qubits are shielded in enclosed boxes \cite{Paik2011}. 
Here we have directly demonstrated that the effects
of cross-coupling on fidelity can be made small for planar integrated
circuits.

%\section{9. CZ}

Control over the interaction strength with nanosecond resolution provides
a unique tool for constructing fast two-qubit gates. In Fig.\,\ref{fig:twoQubits}(a)
we illustrate a method for using dynamic coupling to implement a fast
controlled-Z (CZ) gate, which has minimal non-adiabatic leakage errors.
The straight lines correspond to the energies of the $|11\rangle$
and $|02\rangle$ states of the uncoupled system. Turning on the interaction
pushes the energy levels apart, with the energies of the coupled system
plotted as curved lines. Adiabatically turning on and off the coupling,
as depicted with arrows, causes the $|11\rangle$ eigenstate to accumulate
a dynamic phase. By calibrating the length of the interaction the
phase shift can be set to $\pi$ for a CZ gate.

In Fig.\,\ref{fig:twoQubits}(b) we use a Ramsey measurement to verify
that the gate sequence produces the desired results. We first apply
a $\pi/2$ pulse to $Q_{1}$, perform a CZ, apply a second $\pi/2$
pulse with varying phase, and then measure the qubit excited state
probability. We then repeat the experiment with an excitation in $Q_{2}$
and overlay the data. The solid lines are fits to cosine oscillations
with zero and $\pi$ phase shifts. The $\pi$ phase shift is observed
only when there is an excitation in each qubit, otherwise the phase
accumulation is zero.

We extract the fidelity of this CZ gate using interleaved randomized
benchmarking, in which we insert a CZ between random single-qubit
Pauli gates. A reference curve without the interleaved CZ is measured
and plotted in Fig.\ref{fig:twoQubits}(c) along with the interleaved
sequence fidelity. Fitting these two curves allows us to extract an
average CZ gate fidelity of $99.07\%$. The dominant error ($\sim0.66\%$)
comes from decoherence, measured by interleaved randomized benchmarking
on the two qubit idle gate (seeRef.\,\cite{Supplement}). This error
can be suppressed by incorporating optimized capacitor geometry and
improved film growth conditions. Surprisingly, despite the short
gate time, the non-adiabatic error resulting from leakage to the $|02\rangle$
state is small ($\sim0.25\%$), measured by the Ramsey error filter
technique (seeRef.\,\cite{Supplement}) \cite{Lucero2008}. This is
the result of an optimized adiabatic trajectory based on a theory
of optimal window functions \cite{Martinis2014}. The adiabatic trajectory
used to vary the coupling strength is shown in the inset in panel
(c).

High-fidelity gates have been demonstrated using Xmon qubits \cite{Barends2014}.
We believe that these fidelities can be further improved by utilizing
the added functionality provided by tunable coupling. This will require
the incorporation of lower loss materials, optimized capacitor geometry
and characterization using the full Clifford group; this is currently
in progress.

In conclusion, we have demonstrated a superconducting qubit architecture
with high coherence and tunable coupling. We explore two distinct
advantages of this architecture for quantum computation. First, the
ability to isolate individual qubits allows for high fidelity local
operations that are not degraded by the presence of neighboring qubits.
Second, by dynamically tuning the interaction strength, we demonstrate
a new two-qubit CZ gate, at a speed approaching that of single-qubit
gates. Combining these features with high coherence, the introduced
architecture represents a viable platform for implementing future
quantum algorithms.

This work was supported by the Office of the Director of National
Intelligence (ODNI), Intelligence Advanced Research Projects Activity
(IARPA), through the Army Research Office grant W911NF-10-1-0334.
C.N., P.R. and M.R.G acknowledge the support form the US National Science Foundation under CDI grant DMR-1029764.
All statements of fact, opinion or conclusions contained herein are
those of the authors and should not be construed as representing the
official views or policies of IARPA, the ODNI, or the U.S. Government.
Devices were made at the UC Santa Barbara Nanofabrication Facility,
a part of the NSF-funded National Nanotechnology Infrastructure Network,
and at the NanoStructures Cleanroom Facility

\bibliographystyle{apsrev4-1}

\end{document}

%%%  Title, date and abstract %%%

\preprint{}

\title{Supplementary information for \textquotedbl{}Qubit architecture with
high coherence and fast tunable coupling\textquotedbl{}}

\author{Yu Chen$^{1}$}

\thanks{These authors contributed equally to this work}

\author{C. Neill$^{1}$}

\thanks{These authors contributed equally to this work}

\author{P. Roushan$^{1}$}

\thanks{These authors contributed equally to this work}

\author{N. Leung$^{1}$}

\author{M.l Fang$^{1}$}

\author{R. Barends$^{1}$}

\author{B. Campbell$^{1}$}

\author{Z. Chen$^{1}$}

\author{B. Chiaro$^{1}$}

\author{A. Dunsworth$^{1}$}

\author{E. Jeffrey$^{1}$}

\author{J. Kelly$^{1}$}

\author{A. Megrant$^{1}$}

\author{J. Y. Mutus$^{1}$}

\author{P. J. J. O'Malley$^{1}$}

\author{C. M. Quintana$^{1}$}

\author{D. Sank$^{1}$}

\author{A. Vainsencher$^{1}$}

\author{J. Wenner$^{1}$}

\author{T. C. White$^{1}$}

\author{Michael R. Geller$^{2}$}

\author{A. N. Cleland$^{1}$}

\author{John M. Martinis$^{1}$}

\email{martinis@physics.ucsb.edu}

\affiliation{$^{1}$Department of Physics, University of California, Santa Barbara,
California 93106-9530, USA}

\affiliation{$^{2}$Department of Physics and Astronomy, University of Georgia,
Athens, Georgia 30602, USA}

\maketitle

\section{Calibration}

A key aspect of our design is the independent control of the qubit
frequency and inter-qubit coupling. The resonance frequency of the
individual qubits depends on the impedance of the coupling circuit;
this is true for any coupling scheme. In our design, the total qubit
inductance $L$ is given by 
\begin{align}
L & =L_{J}+L_{g}||(L_{g}+L_{c})\nonumber \\
 & =L_{J}+L_{g}-M
\end{align}
where || stands for \textquotedbl{}in parallel with\textquotedbl{}
and M is the mutual inductance given in Eq.\,(1) of the main text.
Changing the inter-qubit coupling is achieved by changing the mutual
inductance, which additionally shifts the qubit's resonance frequency.
We are able to compensate for this change in inductance using the
tunable inductance of the qubit junction $L_{J}$. The compensation
is achieved by first measuring the qubit frequency $\omega$ as a
function of the qubit flux bias $\Phi_{Q}$ and then as a function
of coupler bias $\Phi_{C}$. The qubit frequency is given by $\omega=1/\sqrt{LC}-\alpha$
where C is the qubit capacitance and $\alpha$ is the anharmonicity.
Solving this expression for L and using the measured data for $\omega$
yields $L(\Phi_{C})$ and $L(\Phi_{Q})$. From the first expression
we determine the change in inductance $\Delta L$ due to a change
in $\Phi_{C}$. Using the second expression we calculate the qubit
flux bias required to shift $L$ by $-\Delta L$. Summing these two
terms yields zero net change in the qubit inductance. Note that the
number of measurements required to compensate for the frequency shift
scales linearly with the number of qubits and couplers.

\begin{figure}
\includegraphics{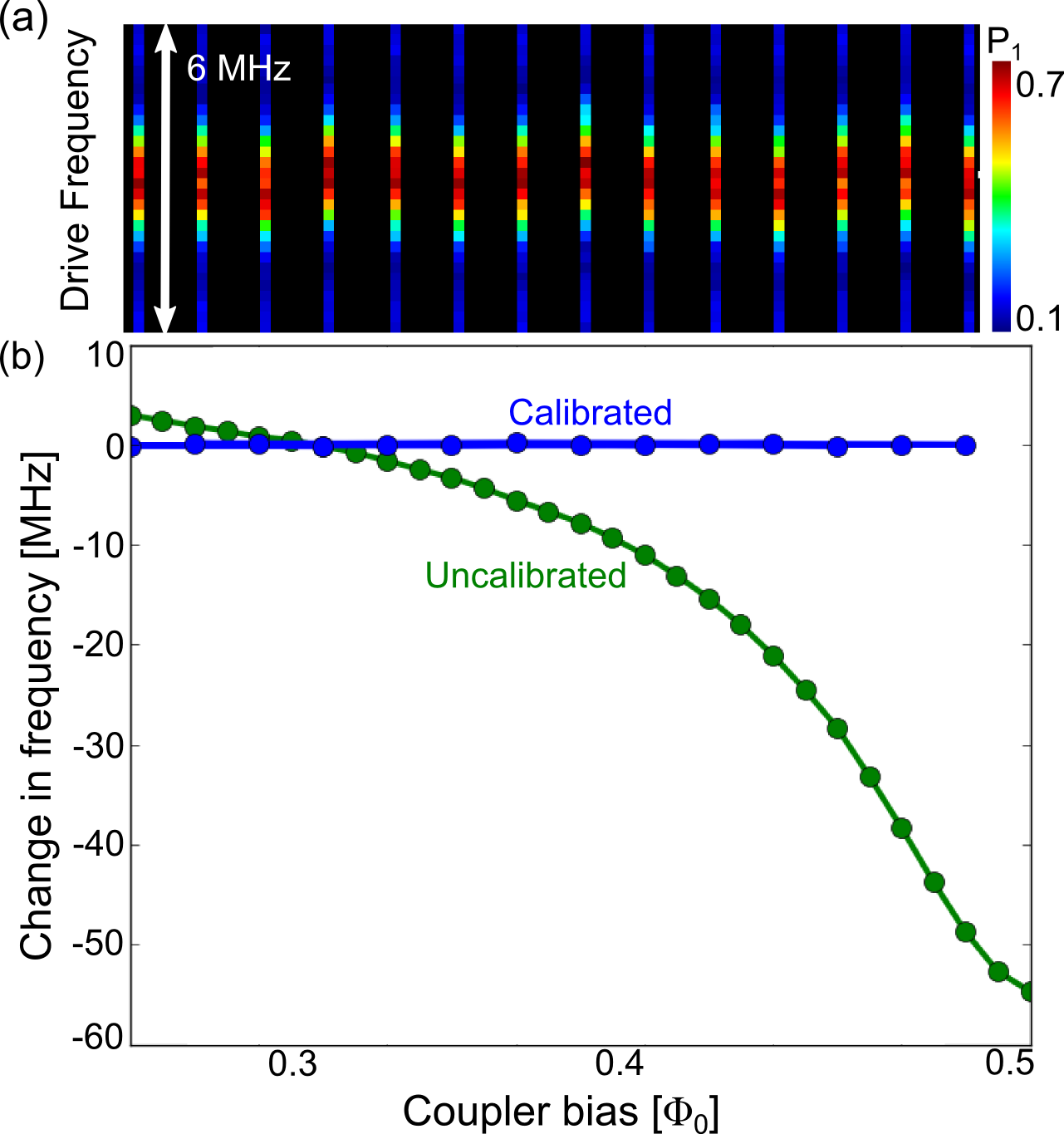} \caption{(a) The frequency of $Q_{1}$, as a function of the coupler flux bias
while the second qubit is far detuned. For each value of the coupling
strength, we compensate the frequency shift due to the change in inductance,
sweep the microwave drive frequency and measure the qubit excited
state probability $P_{1}$. Each line is fit for a peak, with the
results plotted in panel (b) in blue. The associated standard deviation
is 110\,kHz. The same experiment is performed without the
calibration and overlayed in green. }

\label{fig:calibration} 
\end{figure}

The results of this compensation protocol are shown in Fig.\,\ref{fig:calibration}(a).
For each value of the coupler flux bias, we sweep the microwave drive
frequency and measure the excited state probability $P_{1}$. The
frequency is almost completely independent of the coupler bias, with
a standard deviation of $110$ kHz. We fit each vertical column of
data for a peak and plot the results in blue in Fig.\,\ref{fig:calibration}(b).
We perform an identical measurement without calibration and overlay the results
in green. We see that the qubit frequency shifts by over 60\,MHz
($\sim$ $g/2\pi$) as we vary the coupler bias.

\section{Coherence}

\begin{figure}
\includegraphics{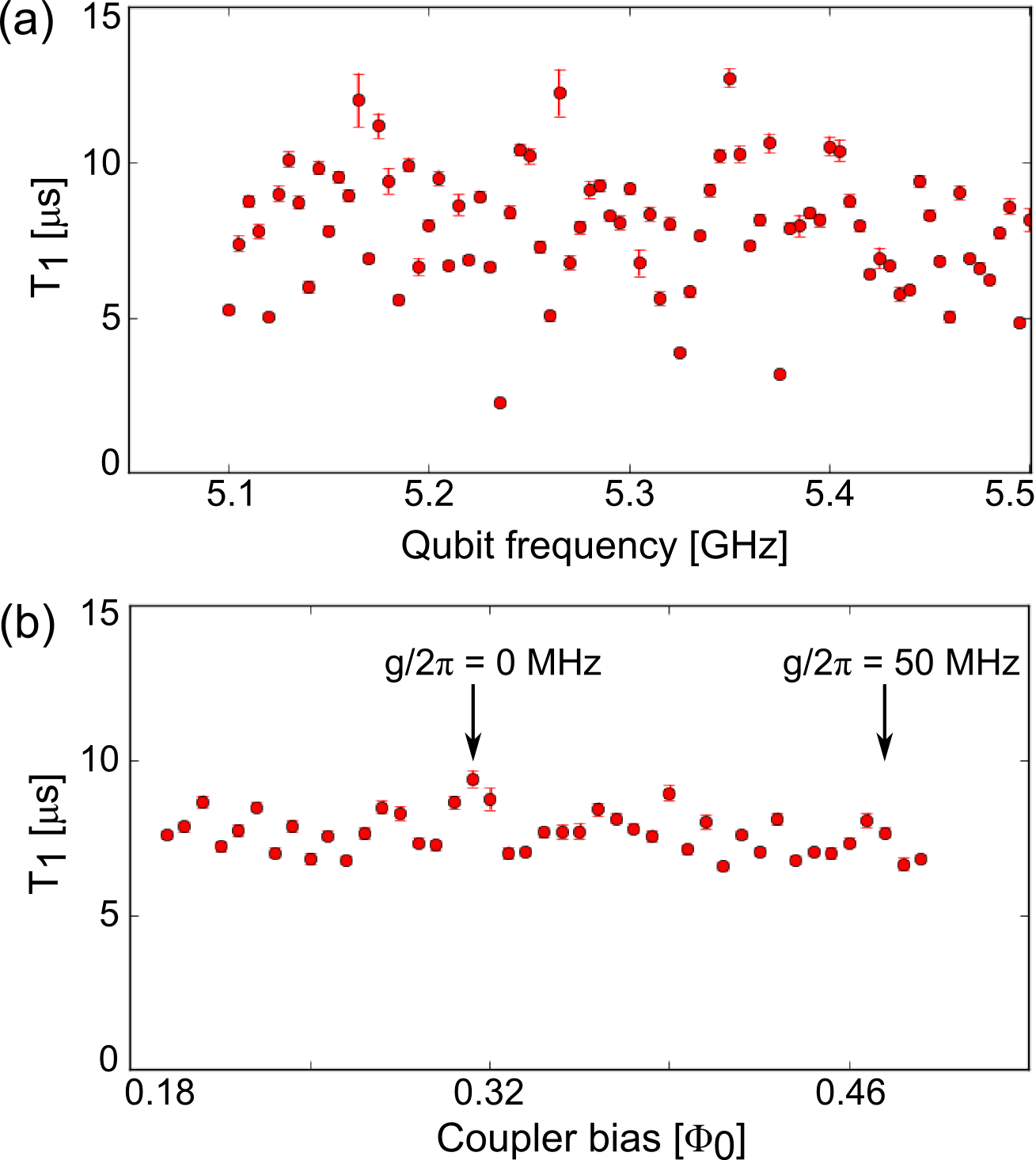} \caption{(a) $T_{1}$ of $Q_1$ as a function of the qubit frequency, when $g = 0$.  These results are comparable to that of the Xmon with similar capacitor geometry and growth conditions. (b) $T_{1}$ of $Q_1$ as a function of the coupler bias, when the qubit frequency is set to $5.3\,$GHz. We find no dependence of the $T_{1}$ on the coupling strength.}

\label{fig:T1} 
\end{figure}

The most important part of constructing this tunable coupling architecture
is to maintain the coherence inherent in the Xmon design. There are
two primary sources of loss associated with the modifications that
we have made: capacitive coupling to surface defects on the coupling
structure and inductive coupling to the added bias line. The voltage
divider created by $L_{J}$ and $L_{g}$ reduce capacitive losses
by a factor of over 2000. 
The coupler bias line has a mutual inductance to the junction loop of 1\,pH; this 1\,pH coupling to a 50\,Ohm line introduces a decoherence source with an associated $T_{1}$ of greater than 200\,$\mu$s at 80\,MHz of coupling. 
We measure $T_{1}$ as a function of the qubit frequency and plot the results in Fig.\,\ref{fig:T1}(a).
These rsults are comparable to the performance of previous Xmon devices with similar capacitor geometry and growth conditions.
We observe no indication that the $T_{1}$ is reduced as we vary the coupling strength, with data shown in Fig.\,\ref{fig:T1}(b).

\section{Zero coupling}

\begin{figure}
\includegraphics{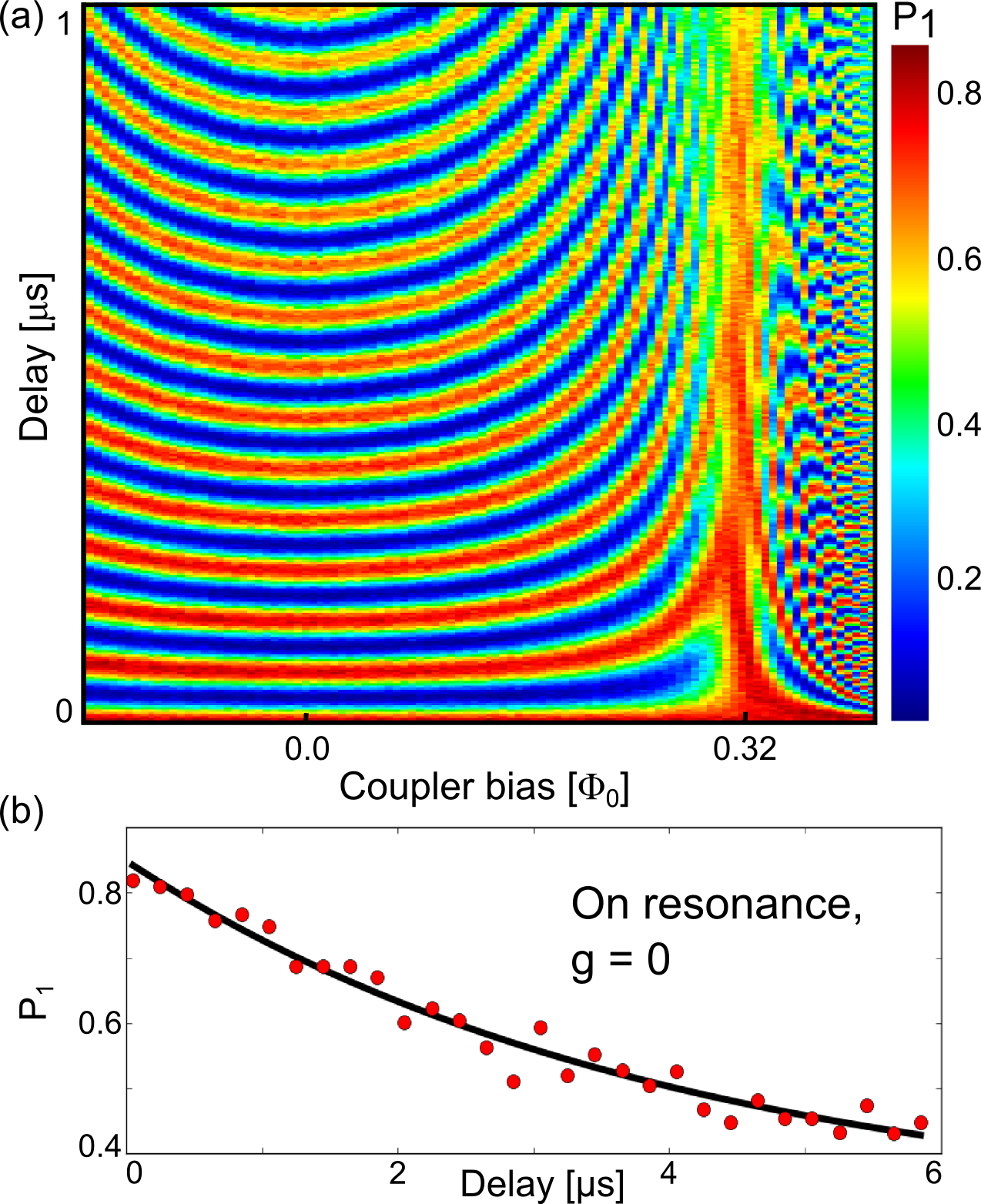} \caption{ (a) Swap spectroscopy for $Q_{1}$, as a function of the coupler
flux bias, with the two qubits on resonance. For each value of the
coupling strength, we excite $Q_{1}$, wait a variable delay time
and measure the excited state probability $P_{1}$. We see no excitation
swapping between the two qubits when coupler bias is $\sim0.32\Phi_0$,
indicating that the coupling is turned off. (b) We set the coupler
bias to this value and examine the excited state probability $P_{1}$
of $Q_{1}$ over an extended delay time. We see no indication of swapping
between the two qubits after 6\,$\mu$s (placing an upper bound on
residual coupling of 50\,kHz.) }

\label{fig:off} 
\end{figure}

An important application of tunable coupling is to isolate individual qubits for local operations by turning off the coupling. 
We characterize the zero coupling of our architecture using a modified swap spectroscopy measurement. 
We bring the two qubits on resonance and vary the coupler flux bias. For each value of the coupling strength, we excite $Q_1$, wait a variable delay time and measure its excited state probability.
As the results in Fig.\,\ref{fig:off}(a) show, over a wide range
of biases, the two qubits can interact and swap an excitation. At
a coupler bias of $\sim0.32 \Phi_0$, there is no excitation swapping between
the two qubits, indicating that the coupling is turned off. Focusing
on zero coupling, we examine the excited state probability $P_{1}$ of
$Q_1$ over a extended delay time, with the results shown in Fig.\,\ref{fig:off}(b).
We see no indication of swapping between the two qubits after 6\,$\mu$s.
This places an upper bound on residual coupling of 50\,kHz, resulting
in an on/off ratio $>1000$.

\section{CZ error budget}

\begin{figure}
\includegraphics{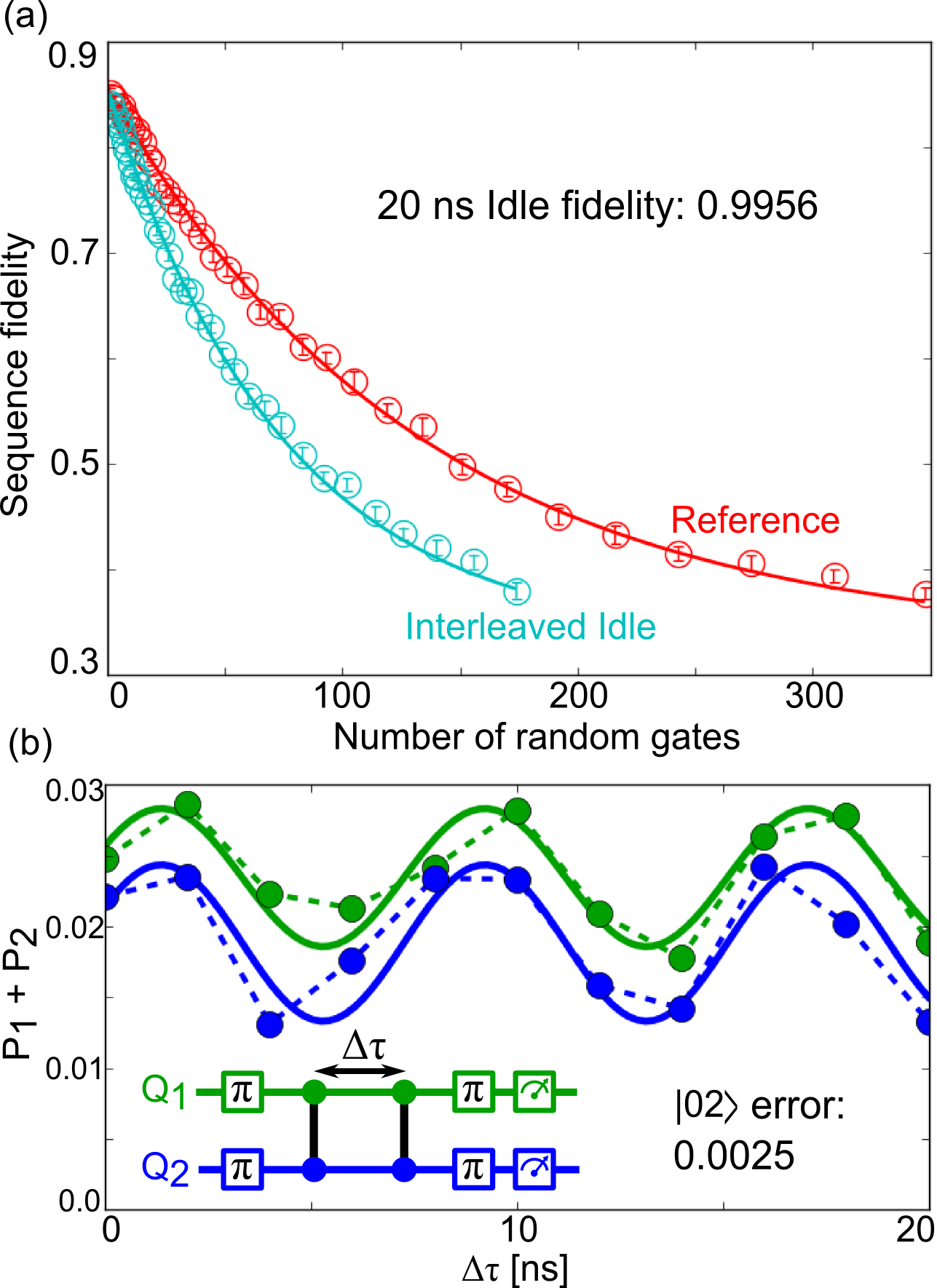} \caption{(a) Interleaved randomized benchmarking on a 20$\,$ns two-qubit idle
gate ($g=0$). We extract a fidelity of $99.56\%$, which suggests
a decoherence error of $0.66\%$ for the 30$\,$ns CZ gate. (b) Inset:
The pulse sequence for the Ramsey error filter technique. Main panel:
The measured excited state probability $P_1 + P_2$ as a function of the delay
between two CZ gates. We observe the expected sinusoidal oscillation
with a peak-to-peak amplitude of $1\%$. The non-adiabatic error from
$|02\rangle$ state leakage is 1/4 of the oscillation amplitude and
is therefore $\thicksim0.25\%$.}

\label{fig:budget} 
\end{figure}

We perform two measurements to determine the sources of errors in
our CZ gate. The dominant contribution to the 0.93\% error comes from decoherence.
We measure this contribution by performing interleaved randomized
benchmarking on a 20\,ns two-qubit idle gate, with $g=0$. We first
measure a reference curve without the interleaved idle and plot the
data in red in Fig.\ref{fig:budget}(a). We then perform an interleaved
randomized benchmarking sequence in which we insert an idle gate between
each random Pauli gate, and overlay the data in blue. Comparing these
two curves allows us to extract a fidelity of $99.56\%$ for a 20$\,$ns
two-qubit idle gate. Scaling this error rate by a factor of 1.5 to
account for the relative length of the CZ yields an error from decoherence
of $\sim0.66\%$.

The next largest contribution to errors are from non-adiabatic transitions
from the $|11\rangle$ to $|02\rangle$ state. We directly measure
this transition using a Ramsey error filter technique \cite{Lucero2008};
the pulse sequence is shown inset in Fig.\,\ref{fig:budget}(b).
We initialize the system in the $|11\rangle$ state and then apply
two CZ gates separated by a variable delay time. Afer applying a $\pi$-pulse
to each qubit, we measure the uncorrelated excited state probability
for each qubit. The results are shown in Fig.\,\ref{fig:budget}(b),
where we see the expected oscillations that result from the interference
between two CZ gates.  The frequency of the oscillation is set by the detuning of the
 $|11\rangle$ and $|02\rangle$ states which was 130 MHz, corresponding to a period of 8 ns. 
The $|02\rangle$ state leakage error is given
as 1/4 of the oscillation amplitude (peak-to-peak). For our 30$\,$ns
CZ gate, we measured a non-adiabatic error of $\sim0.25\%$. This
is suprisingly small considering such a short gate time, and can be
exponentially surpressed with increasing gate length.

\section{Transmon physics}

The operation of the transmon has been previously described in detail
\cite{Koch2007}. Here, we give a simplified calculation in the phase
basis that is useful to describe more complex transmon circuits, as
for the gmon architecture.

Since the transmon produces qubit behavior from a weak non-linearity,
we first review the physics of a linear inductor-capacitor $(LC)$
oscillator. In terms of physical variables charge $q$ and flux $\Phi$,
the oscillator Hamiltonian is given by 
\begin{align}
\hat{H}_{o}=\frac{\hat{q}^{2}}{2C}+\frac{\hat{\Phi}^{2}}{2L}\ .
\end{align}
Here the quantum operators of flux and charge obey the standard commutation
relation $[\hat{\Phi},\hat{q}]=i\hbar$. The oscillator frequency
is the classical value $\omega=1/\sqrt{LC}$, and eigenstates $m$
have energy $E_{m}=\hbar\omega(m+1/2)$. The ground state wavefunction
is given by 
\begin{align}
\Psi_{0}(\Phi)\propto\exp[-(\omega C/2\hbar)\Phi^{2}]\ .\label{eq:width}
\end{align}
Note that the width of the wavefunction is set by the oscillator impedance
$Z_{o}=1/\omega C=\omega L=\sqrt{L/C}$. Varying this impedance changes
the widths of the charge and flux wavefunctions, as illustrated in
Table\,\ref{tab:Zo}. The impedance is also important since it is
used to describe how strongly the oscillator couples to other modes.
The flux and charge operators are conveniently expressed in terms
of the raising and lowering operators 
\begin{align}
\hat{\Phi} & =(\hbar/2\omega C)^{1/2}(a^{\dagger}+a)\\
\hat{q} & =\big(\hbar\omega C/2\big)^{1/2}i(a^{\dagger}-a)\ .
\end{align}

\begin{table}[b]
\caption{\label{tab:Zo} Table of relative width of charge and flux wavefunctions
as capacitance $C$ (and impedance $Z_{o}$) are changed.}

\begin{tabular}{|c|c|c|c|}
\hline 
$C$  & $Z_{o}$  & $\langle\hat{q}^{2}\rangle$  & $\langle\hat{\Phi}^{2}\rangle$ \tabularnewline
\hline 
\ \ small \ \  & \ \ large \ \  & \ \ small \ \  & \ \ large \ \ \tabularnewline
large  & small  & large  & small \tabularnewline
\hline 
\end{tabular}
\end{table}

For a tunnel junction with shunting capacitor, the charge on the metal
island takes on discrete values corresponding to the number of Cooper
pairs $n$. The Hamiltonian for this system is given by 
\begin{align}
\hat{H}_{t}=4E_{c}(\hat{n}-n_{g})^{2}-E_{J}\cos\hat{\delta}\ ,\label{eq:Ht}
\end{align}
where $E_{c}=e^{2}/2C$ is the charging energy and $E_{J}=I_{0}\Phi_{0}/2\pi$
is the Josephson energy from the tunnel junction, with critical current
$I_{0}$. The normalized coordinates are related to ordinary electrical
variables by $\hat{q}=2e\hat{n}$ and $\hat{\Phi}=(\Phi_{0}/2\pi)\hat{\delta}$,
and thus their commutation relation is $[\hat{\delta},\hat{n}]=i$.
Here we have included a \textit{continuous} charge bias $n_{g}$,
produced for example by a small coupling capacitor with voltage bias.
The Josephson term can be written as $\cos\hat{\delta}=[\exp(+i\hat{\delta})+\exp(-i\hat{\delta})]/2$,
corresponding to number displacement operators $\exp(\pm i\hat{\delta})$
that couple states that differ by one in the number of Cooper pairs.

The form of the solution for this Hamiltonian depends on the ratio
of these two energies. For small capacitance where $E_{c}\gg E_{J}$,
the ``Cooper-pair box'' limit, the charging energy dominates, and
the eigenstates are described by one or the superposition of two number
states. The states sensitively depend on the gate charge $n_{g}$.
This is death to qubit physics, since fluctuations of gate charge
from the movement of trapped charge around the junction produces large
qubit decoherence from dephasing.

We are interested in the large capacitance ``transmon'' limit, where
$E_{J}\gg E_{c}$. Here, the dependence of qubit energy on the gate
charge becomes exponentially small, so qubit decoherence from charge
fluctuations essentially vanishes. To understand this, note that for
large capacitance the phase fluctuations are small. The potential
$\cos\hat{\delta}$ can then be expanded in powers of $\hat{\delta}$,
with the lowest non-trivial term giving an inductive energy. First
considering the case $n_{g}=0$, one obtains a harmonic oscillator-like
Hamiltonian 
\begin{align}
H_{to}=4E_{c}\hat{n}^{2}+(\Phi_{0}/2\pi)^{2}\hat{\delta}^{2}/2L_{J}\ ,
\end{align}
where the Josephson inductance is $L_{J}=(\Phi_{0}/2\pi)^{2}/E_{J}=\Phi_{0}/2\pi I_{0}$.
We can thus use harmonic oscillator solutions as the basis eigenstates
for perturbation theory.

Note that formally the charge wavefunction is a delta-function comb
with spacings $2e$ in charge, with amplitudes given by the harmonic
oscillator solution. The charge comb corresponds to a phase wavefunction
periodic in $2\pi$. As the capacitance increases, the number of states
in the charge wavefunction increases, so that the relative separation
of the teeth in the charge comb become so closely spaced as to look
like the normal \textit{continuous} solution for the harmonic oscillator.
In phase, this implies the wavefunction is so localized in phase that
the $2\pi$ periodicity does not matter.

The phase wavefunction has a width $\langle\hat{\delta}^{2}\rangle$
that can be computed using the exponential term in the wavefunction
given by Eq.\,(\ref{eq:width}) 
\begin{align}
1 & =\frac{\omega C}{\hbar}\Big(\frac{\Phi_{0}}{2\pi}\Big)^{2}\langle\hat{\delta}^{2}\rangle\ ,
\end{align}
which gives 
\begin{align}
\langle\hat{\delta}^{2}\rangle & =\sqrt{8E_{c}/E_{J}}\\
 & =Z_{J}/(R_{K}/8\pi)\ ,
\end{align}
where in the last equation $R_{K}=h/e^{2}=25.8\,\textrm{k}\Omega$
is the resistance quantum, and $R_{K}/8\pi=1.026\,\textrm{k}\Omega$.
The phase basis works well when the mean quantum fluctuation of the
phase is small, which corresponds to a small $E_{c}/E_{J}$ ratio
or a junction impedance $Z_{J}=\sqrt{L_{J}/C}$ much less than $1\,\textrm{k}\Omega$.

The effect of the gate charge $n_{g}$ in the Hamiltonian can be computed
by noting that this offset in the operator $\hat{n}$ can be accounted
for by the displacement operator $\exp(in_{g}\hat{\delta})$ applied
to the solution of $H_{t}$ with $n_{g}=0$. This is equivalent to
imposing periodic boundary conditions at the phase $\delta=\pm\pi$
\begin{align}
\Psi(-\pi)=\Psi(\pi)\ e^{i2\pi n_{g}}\ .
\end{align}
We can estimate the effect of this boundary condition on the eigenstates
by noting that it should be proportional to the probability of the
wavefunction at $\delta=\pi$. Using the harmonic oscillator solution,
the magnitude of the modulation of eigenstate energy from charge $n_{g}$
should scale approximately as 
\begin{align}
\Delta E & \propto|\Psi_{0}(\delta=\pi)|^{2}\\
 & =\exp[-(\omega C/\hbar)(\Phi_{0}/2)^{2}]\\
 & =\exp[-(\pi^{2}/8)\sqrt{8E_{J}/E_{c}}]\ .
\end{align}

We may calculate the exponential factor precisely by including the
non-linear junction energy. Using the WKB theory, with constants $2m=1/4E_{c}$
and $\hbar=1$ from Eq.\,(\ref{eq:Ht}) and its commutation relation,
we find 
\begin{align}
|\Psi_{0}(\pi)|^{2} & =\exp[-2\int_{0}^{\pi}d\delta\sqrt{(1/4E_{c})E_{J}(1-\cos\delta)}\,]\\
 & =\exp[-\sqrt{8E_{J}/E_{c}}]\ ,
\end{align}
matching the result of Ref.\,\cite{Koch2007}. A large
$E_{J}/E_{c}$ ratio gives exponentially low sensitivity to charge
noise.

Note that the phase qubit has vanishing sensitivity to charge noise
for two reasons. First, the ratio of $E_{J}/E_{c}$ is even larger
than for the transmon. Second, the latest versions of the device used
a shunting inductor for current biasing. The continuous flow of charge
across the junction then shunts any DC change in charge bias. This
latter effect is the purpose of the inductor shunt in the fluxonium
device.

For completeness, we compute the change in the harmonic oscillator
energy eigenvalues due to the cosine nonlinearity. Starting from 
\begin{align}
\cos\hat{\delta}\simeq1-\hat{\delta}^{2}/2+\hat{\delta}^{4}/24\ ,
\end{align}
the correction to the energy from the fourth order term is 
\begin{align}
\Delta E_{m} & =-E_{J}\langle m|\hat{\delta}^{4}|m\rangle/24\\
 & =-\frac{E_{J}}{24}\Big(\frac{\hbar}{2\omega C}\Big)^{2}\Big(\frac{2\pi}{\Phi_{0}}\Big)^{4}\langle m|(a^{\dagger}+a)^{4}|m\rangle\ .\label{eq:em}
\end{align}
The matrix element can be calculated by using the square $(a^{\dagger}+a)^{2}=a^{\dagger2}+a^{2}+2a^{\dagger}a+1$,
giving 
\begin{align}
\langle m|(a^{\dagger} & +a)^{4}|m\rangle=\langle m|a^{\dagger2}a^{2}+a^{2}a^{\dagger2}+(2a^{\dagger}a+1)^{2}|m\rangle\\
 & =m(m-1)+(m+1)(m+2)+(2m+1)^{2}\\
 & =6m^{2}+6m+3\,
\end{align}
where in the first equation we have only kept terms that leave $|m\rangle$
unchanged. The change in energy between adjacent states is 
\begin{align}
\Delta(E_{m}-E_{m-1})=-mE_{c}
\end{align}
as expected. As the unperturbed oscillator frequency can be written
as $\hbar\omega=\sqrt{8E_{J}E_{c}}$, the fractional change in qubit
frequency is $\sqrt{E_{c}/8E_{J}}$.

\subsection{Series Inductance}

We next consider how this physics changes when including an inductance
$L$ in series with the Josephson junction. The total phase across
the two elements is given by $\delta=\delta_{L}+\delta_{J}$. The
conservation of current at the node between the two elements gives
the constraint $I_{L}=I_{0}\sin\delta_{J}$, which then can be used
to relate the individual phase changes and their derivative 
\begin{align}
\delta_{L}/L & =\sin\delta_{J}/L_{J0}\\
d\delta_{L}/L & =d\delta_{J}\cos\delta_{J}/L_{J0}\ ,
\end{align}
where we have defined $L_{J0}=\Phi_{0}/2\pi I_{0}=(\Phi_{0}/2\pi)^{2}/E_{J}$
as the Josephson inductance at zero current.

The WKB theory gives a charge sensitivity that includes both Josephson
and inductor energies 
\begin{align}
- & \ln|\Psi_{0}(\pi)|^{2}\nonumber \\
= & \sqrt{\frac{1}{E_{c}}}\int_{0}^{\pi}d\delta\sqrt{E_{J}(1-\cos\delta_{J})+(\delta_{L}\Phi_{0}/2\pi)^{2}/2L}\\
= & \sqrt{\frac{E_{J}}{E_{c}}}\int_{0}^{\pi}d\delta_{J}[1+(L/L_{J0})\cos\delta_{J}]\nonumber \\
 & \ \ \ \ \ \ \ \ \ \ \times\sqrt{1-\cos\delta_{J}+(L/2L_{J0})\sin^{2}\delta_{J}}\\
\simeq & \sqrt{8E_{J}/E_{c}}\,(1-0.166\, L/L_{J0})\ ,\label{eq:lin}
\end{align}
where the integral was evaluated numerically. The linear expansion
in Eq.\,(\ref{eq:lin}) is quite good for $L/L_{J0}\le1$

The nonlinearity in the energy levels can be evaluated by noting that
the quantum fluctuations of the phase is small, so that we can use
the linear relation for phase change $\delta_{L}/L=\delta_{J}/L_{J0}$.
The junction phase can then be found using an inductance divider relation
\begin{align}
\delta_{J}=\frac{L_{J0}}{L+L_{J0}}\delta\ .
\end{align}
Following Eq.\,(\ref{eq:em}), the change in energy eigenvalues is
proportional to $\langle\hat{\delta}_{J}^{4}\rangle=\langle\hat{\delta}^{4}\rangle/(1+L/L_{J0})^{4}$,
giving 
\begin{align}
\Delta(E_{m}-E_{m-1}) & =-\frac{E_{J}}{24}\Big(\frac{\hbar}{2\omega C}\Big)^{2}\Big(\frac{2\pi}{\Phi_{0}}\Big)^{4}\frac{12m}{(1+L/L_{J0})^{4}}\\
 & =-mE_{c}\frac{1}{\omega^{2}L_{J0}C}\frac{1}{(1+L/L_{J0})^{4}}\\
 & =-mE_{c}\frac{1}{(1+L/L_{J0})^{3}}\ ,
\end{align}
where for the last equation we have used the resonance condition $\omega^{2}=1/(L+L_{J0})C$.
We see that the extra linear inductance lowers the nonlinearity coming
from the junction.